\documentclass[10pt,aps,prb,twocolumn,preprintnumbers,amsmath,amssymb,floatfix,showpacs,citeautoscript]{revtex4-1}
\usepackage[latin1]{inputenc}
\usepackage[T1]{fontenc}
\usepackage[english]{babel}
\usepackage[pdftex]{graphicx}
\usepackage{amsmath}
\usepackage{times}
\usepackage[usenames,dvipsnames,svgnames,table]{xcolor}
\usepackage[colorlinks,plainpages=false,linkcolor=blue,urlcolor=blue,citecolor=blue,pdfpagemode=UseNone]{hyperref}

\renewcommand{\AA}{\text{\r{A}}}

\newcommand\Vek[1]{\vec{#1}}

\newcommand\muB{\mu_{\text{B}}}

\begin{document}

\title
{
\boldmath
Competition of defect ordering and site disproportionation in strained LaCoO$_{3}$ on SrTiO$_3$(001)
}

\author{Benjamin Geisler}
\email{benjamin.geisler@uni-due.de}
\affiliation{Fakult\"at f\"ur Physik, Universit\"at Duisburg-Essen and Center for Nanointegration (CENIDE), Campus Duisburg, Lotharstr.~1, 47048 Duisburg, Germany}
\author{Rossitza Pentcheva}
\email{rossitza.pentcheva@uni-due.de}
\affiliation{Fakult\"at f\"ur Physik, Universit\"at Duisburg-Essen and Center for Nanointegration (CENIDE), Campus Duisburg, Lotharstr.~1, 47048 Duisburg, Germany}

\date{\today}

\begin{abstract}
The origin of the $3 \times 1$ reconstruction observed in epitaxial LaCoO$_{3}$ films on SrTiO$_3(001)$
is assessed by using first-principles calculations including a Coulomb repulsion term.
We compile a phase diagram as a function of the oxygen pressure,
which shows that ($3 \times 1$)-ordered oxygen vacancies (LaCoO$_{2.67}$) are favored under commonly used growth conditions, while stoichiometric films emerge under oxygen-rich conditions.
Growth of further reduced LaCoO$_{2.5}$ brownmillerite films is impeded by phase separation.
We report two competing ground-state candidates for stoichiometric films:
a semimetallic phase with $3 \times 1$ low-spin/intermediate-spin/intermediate-spin magnetic order and a semiconducting phase with intermediate-spin magnetic order.
This demonstrates that tensile strain induces ferromagnetism even in the absence of oxygen vacancies.
Both phases exhibit an intriguing ($3 \times 1$)-reconstructed octahedral rotation pattern and accordingly modulated La-La distances.
In particular, charge and bond disproportionation and concomitant orbital order of the $t_{2g}$ hole emerge at the Co sites that are also observed for unstrained bulk LaCoO$_3$ in the intermediate-spin state and explain structural data obtained by x-ray diffraction at elevated temperature.
Site disproportionation drives a metal-to-semiconductor transition that reconciles the
intermediate-spin state with the experimentally observed low conductivity during spin-state crossover without Jahn-Teller distortions.
\end{abstract}


\maketitle

\section{Introduction}

Lanthanum cobaltate (LaCoO$_3$, LCO) is a correlated transition metal oxide that shows rich and intriguing physics related to spin-state crossover:
Since Co$^{3+}$ is in a $d^6$ configuration, the ground state is nonmagnetic (NM) and insulating  with fully occupied $t_{2g}$ states (low spin, LS)
up to a temperature of $\sim 50$~K.~\cite{YanZhouGoodenough:04, Klie:07, Doi:14}
Above $100$~K, LCO is a paramagnetic semiconductor
that undergoes a transition to a metal between $400$ and $600$~K.~\cite{YanZhouGoodenough:04, Klie:07, Tachibana-HeatCapacityLCO:08, Doi:14}
The mechanism behind this behavior,
particularly whether an intermediate-spin (IS) state~\cite{Korotin:96, Maris-LCO-OrbitalOrder:03, Ishikawa-LCO:04, Klie:07, RondinelliSpaldin:09} or a high-spin (HS) state~\cite{Haverkort:06, KarolakLichtenstein:15, Krapek:12, Shimizu:17, Sterbinsky:18} is thermally excited
($S=1$ or $S=2$),
or a mixture of both~\cite{Knizek-Ku-LCO:09, Doi:14},
is still controversially debated.~\cite{Ikeda:16}
A key argument against the IS state is its metallic conductivity predicted by first-principles simulations~\cite{Knizek-Ku-LCO:09, HsuLeighton-LCO:10}.

Recently, a $3 \times 1$ reconstruction was observed in epitaxial LCO films on SrTiO$_3(001)$ (STO)
due to tensile strain,
expressed in a striped pattern appearing along the $[100]$ direction
in transmission electron microscopy (TEM) images.~\cite{Choi:12, Kwon:14, BiskupVarela:14, Mehta:15}
These stripes were shown to be related to a short$/$short$/$long modulation of the La-La distances
along the $[100]$ direction.~\cite{BiskupVarela:14}
Moreover, an insulating~\cite{Freeland-LCOSTO:08} and ferromagnetic (FM) ground state emerges,
with a Curie temperature of $T_\text{C} \approx 80$~K.~\cite{Fuchs:08, Freeland-LCOSTO:08, Mehta:09, Choi:12, BiskupVarela:14, Mehta:15, Qiao-LCO:15, Feng-LCO:19}

Two distinct models have been put forward to explain these observations.
The first model is based on an ordered arrangement of oxygen vacancies in every third Co $(100)$ plane,~\cite{BiskupVarela:14}
formally LaCoO$_{2.67}$.
The released charges are accommodated by
a Peierls-like modulation of the La-La distances
in conjunction with a complex electronic reconstruction of the Co valence state,
resulting in wide-gap insulating films and FM order. 
The second model assumes a LS/LS/HS-modulated magnetic order of successive Co $(100)$ planes
in stoichiometric LCO films.~\cite{Choi:12, Kwon:14}
The La-La distances across the HS planes are expanded
with respect to those across the LS planes.
First-principles calculations revealed that this $3 \times 1$ spin-ordered state
is lower in energy than a bulklike LS/LS/LS NM state for tensile strain,
but were constrained by small supercells and the lack of octahedral rotations.~\cite{Kwon:14}

Here we present a systematic first-principles study of epitaxial LCO films grown on STO$(001)$.
Large supercells provide extensive degrees of freedom
and account for structural, electronic, and magnetic reconstruction mechanisms,
oxygen vacancies, and distinct octahedral rotation patterns,
which we compile in a phase diagram as a function of the oxygen pressure.
Under oxygen-poor conditions that are are typically used during growth,
particularly in those experiments that observed a $3 \times 1$ reconstruction,~\cite{Choi:12, Kwon:14, BiskupVarela:14, Mehta:15}
ordered oxygen vacancies (LaCoO$_{2.67}$) stabilize,
which lead to insulating FM films with a strong La-La distance modulation that agree with experimental observations.~\cite{BiskupVarela:14}
In contrast,
oxygen-rich conditions favor the formation of
stoichiometric films,
for which we report two competing ground-state candidates:
a $3 \times 1$ spin-reconstructed LS/IS/IS phase and a phase of pure IS magnetic order.
Both phases exhibit a peculiar octahedral rotation pattern, small modulations of the La-La distances, and charge and orbital order which can only emerge in large supercells.
While the first phase is semimetallic with reconstructed Fermi surface topology, the second phase is semiconducting due to site disproportionation.
We argue that tensile strain is sufficient to induce FM order,
even in the absence of oxygen vacancies.

Reducing the concentration of oxygen,
we additionally explore the structural, electronic, and magnetic properties of brownmillerite LaCoO$_{2.5}$ films strained on STO$(001)$.
We find that the formation of brownmillerite is impeded due to the tendency to decompose to CoO and La$_2$O$_3$ under the corresponding growth conditions.

We also report a novel IS phase of bulk LCO,
in which the excited electron is accommodated in the $e_{g}$ states via a charge- and bond-disproportionation mechanism,
which bears similarities to nickelate systems~\cite{ABR:11, RENickelateReview:16, WrobelGeisler:18, GeislerPentcheva-LNOLAO:18},
while orbital order occurs exclusively for the hole in the $t_{2g}$ manifold.
Breathing-mode distortions lead to O-Co-O distances that are close to x-ray diffraction results at elevated temperature~\cite{Maris-LCO-OrbitalOrder:03},
whereas Jahn-Teller distortions are found to be minute.
The emerging band gap offers a yet unexplored route to reconcile the otherwise metallic IS state with the experimentally observed low conductivity during the thermally driven spin transition of bulk LCO.~\cite{YanZhouGoodenough:04, Klie:07, Doi:14}

\vspace{-3.0ex}

\section{Computational details}

\vspace{-1.0ex}

\begin{figure}[b]
	\centering
	\includegraphics[]{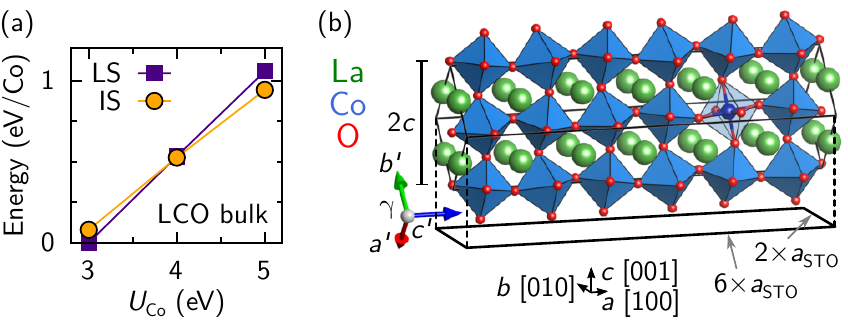}
	\caption{(a)~Evolution of the total energy of bulk LCO in the LS and IS state with $U_\text{Co}$. The LS ground state becomes unfavorable for $U_\text{Co} \gtrsim 4$~eV. (b)~Geometry of the monoclinic supercell used to model LCO films grown on STO$(001)$. The substrate coordinate system is described by ($a,b,c$), whereas ($a',b',c'$) denote the supercell axes.}
	\label{fig:HubbardU-Geometry}
\end{figure}

We performed first-principles calculations in the framework
of spin-polarized density functional theory~\cite{KoSh65} (DFT)
as implemented in the Quantum ESPRESSO code.~\cite{PWSCF}
The generalized gradient approximation was used for the exchange-correlation functional  
as parametrized by Perdew, Burke, and Ernzerhof.~\cite{PeBu96}
Static correlation effects were considered within the DFT$+U$ formalism~\cite{QE-LDA-U:05}.
We confirmed the earlier finding~\cite{RondinelliSpaldin:09}
that the NM LS ground state of bulk LCO is destabilized for $U_\text{Co} \gtrsim 4$~eV [Fig.~\ref{fig:HubbardU-Geometry}(a)].
Thus, we consistently use $U_\text{Co}=3$~eV throughout this work,
in line with previous studies.~\cite{RondinelliSpaldin:09, Knizek-Ku-LCO:09, BiskupVarela:14, Kwon:14, KarolakLichtenstein:15}
Lower values render too small band gaps at variance with experiments~\cite{Abbate:93, Chainani:92}.
We confirmed our main results by using the rotationally invariant Liechtenstein approach~\cite{LiechtensteinAnisimov:95} with $U=4$~eV and $J=1$~eV.

In order to provide sufficient degrees of freedom
for octahedral rotations, structural, electronic, and magnetic reconstructions, and oxygen vacancies,
we modeled LCO films on STO$(001)$ by using monoclinic supercells containing up to $60$ atoms.
From the substrate lattice parameter $a_\text{STO} = 3.905~\AA$
and the pseudocubic cell height $c$,
the supercell geometry [Fig.~\ref{fig:HubbardU-Geometry}(b)] follows as
$\gamma = 2 \arctan (c / a_\text{STO})$,
$a' = b' = a_\text{STO} \sqrt{1 + ( c / a_\text{STO} )^2}$, and
$c' = 6 \, a_\text{STO}$.
Note that due to the antiferrodistortive octahedral rotations, a supercell with $c' = 3 \, a_\text{STO}$ is not sufficient. Experimentally, the $3 \times 1$ reconstruction is determined on the basis of the La-La distances~\cite{BiskupVarela:14},
whereas details of the octahedral rotations have not been explored so far.
For LCO films in brownmillerite structure we used $c' = 4 \, a_\text{STO}$ and $36$ atoms.
The supercells are rotated by $45^\circ$ around the $[100]$~axis [Fig.~\ref{fig:HubbardU-Geometry}(b)]
and the pseudocubic cell height $c$ was optimized in all cases.

Conventionally, bulk LCO is described by a rhombohedral unit cell~\cite{RondinelliSpaldin:09}
to account for the antiferrodistortive octahedral rotations ($R\bar{3}c$ symmetry).
For better comparability with our supercell results for epitaxial films,
we use a $20$-atom $\sim \sqrt{2}a \times \sqrt{2}a \times 2a$ monoclinic unit cell,
thereby providing additional degrees of freedom.
The cell is chosen such that the La-La distances are equal to
the pseudocubic lattice parameter $a_\text{LCO} = 3.83~\AA$,~\cite{RadaelliCheong:02}
which is close to our optimized $3.84~\AA$.

\begin{figure}[b]
	\centering
	\includegraphics[]{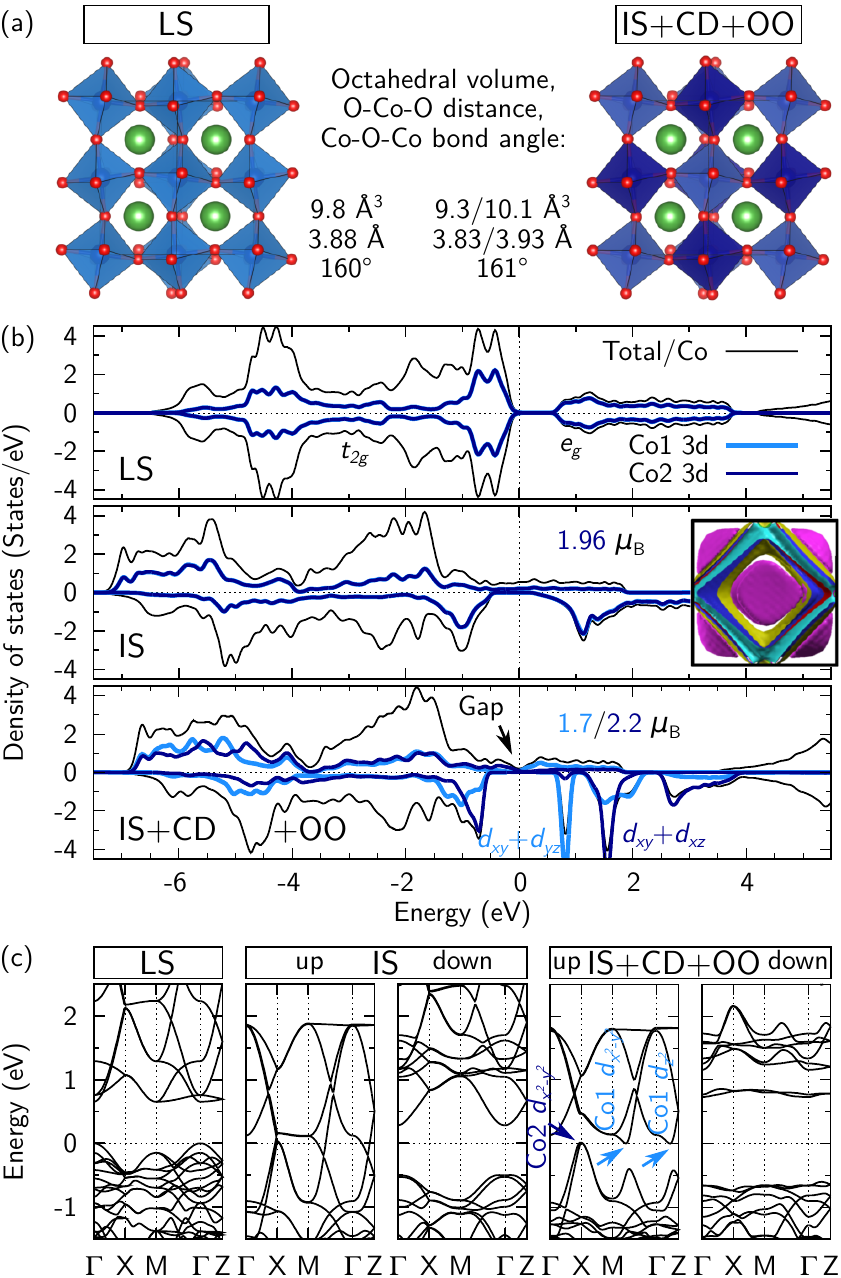}
	\caption{Impact of different spin states in bulk LCO. (a)~Optimized structures of bulk LCO in the LS and the IS+CD+OO state. (b)~Spin-resolved total and projected DOS for the LS, the unstable IS, and the stable IS+CD+OO state (cf.~Fig.~\ref{fig:LCO-Bulk-OO}). The Fermi energy has been chosen as reference. A Fermi surface is provided for the metallic IS system. (c)~Corresponding (spin-resolved) band structures, highlighting the emerging band gap in the IS+CD+OO state. The arrows indicate the distinct band characters at the band edges.}
	\label{fig:LCO-Bulk}
\end{figure}

\begin{figure*}[]
	\centering
	\includegraphics[]{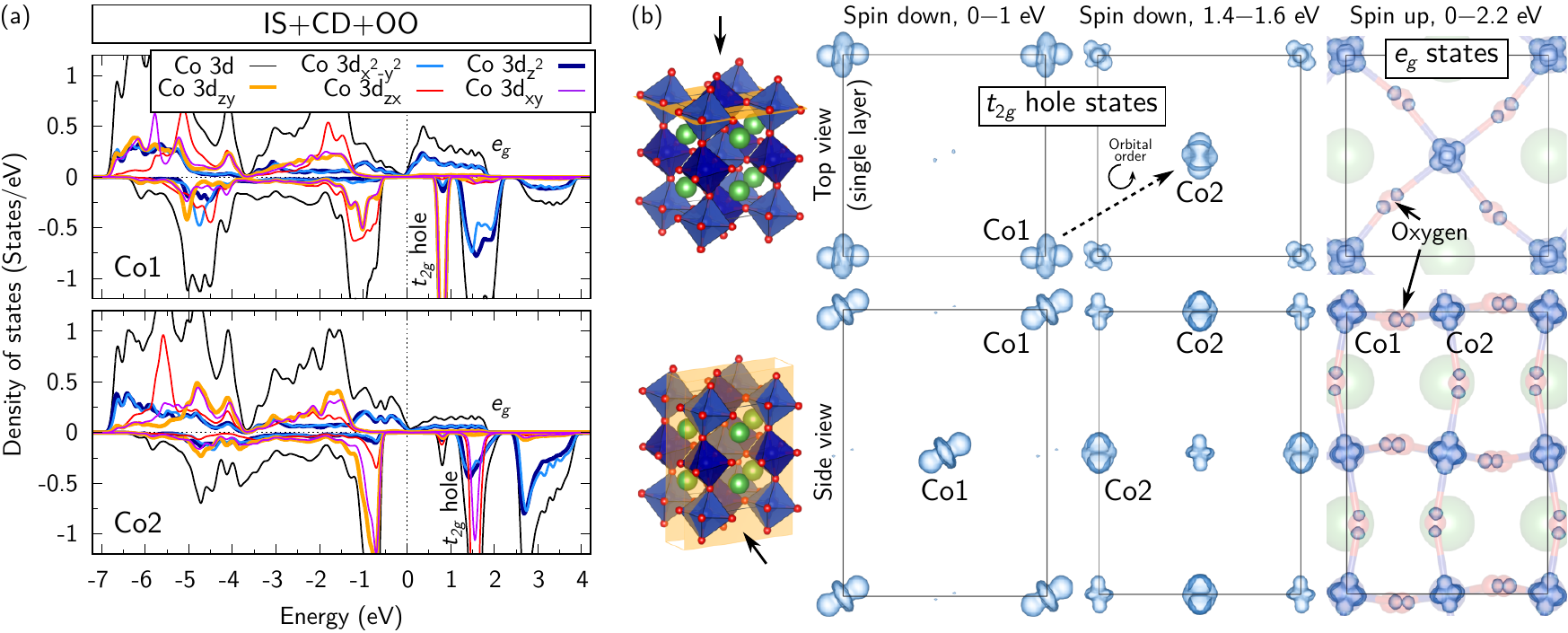}
	\caption{Electronic structure of the IS+CD+OO phase in bulk LCO. (a)~Site-, spin-, and orbital-resolved DOS. (b)~Top and side views of isosurfaces of the integrated local densities of states~\cite{Geisler-TMs:15, GeislerPentcheva-LNOLAO-Resonances:19} (ILDOS) for different spin channels and energy intervals, analyzing the \textit{empty} states. Integration of the sharp resonances at $\sim 0.8$ and $1.55$~eV shows the orbital order in the minority spin channel induced by the hole. The apparent $d_{z^2}$ orbitals are actually superpositions of $t_{2g}$ orbitals and not aligned with any O-Co-O axis; cf.~Ref.~\onlinecite{Cwik-LaTiO3:03}. In contrast, the mapping of the majority spin $e_g$ states underlines the CD, absence of orbital order, and reveals their hybridization with O~$2p$ orbitals.}
	\label{fig:LCO-Bulk-OO}
\end{figure*}

Wave functions and density were expanded into plane waves up to cutoff energies of $35$ and $350$~Ry, respectively.
Ultrasoft pseudopotentials,~\cite{Vanderbilt:1990}
as successfully employed in previous work,~\cite{Suzuki:13, Geisler:13, ComtesseGeisler:14, GeislerPopescu:14, Geisler-Heusler:15, Geisler-LNOSTO:17, WrobelGeisler:18, GeislerPentcheva-LNOLAO:18, Viewpoint:19, GeislerFePcHSi:19}
were used, 
treating the
La $5s$, $5p$, $5d$, $6s$, $6p$,
Co $3d$, $4s$, $4p$,
and O $2s$, $2p$
atomic subshells as valence states.
For La and Co a non-linear core correction~\cite{LoFr82} was included.
Different Monkhorst-Pack $\Vek{k}$-point grids~\cite{MoPa76} were used
together with a Methfessel-Paxton smearing~\cite{MePa89} of $10$~mRy to sample the Brillouin zone:
$16 \times 16 \times 4$ for the large supercell models,
$16 \times 16 \times 6$ for LCO in brownmillerite structure, and
$16 \times 16 \times 8$ for monoclinic LCO bulk.
The atomic positions were optimized until the maximum component of the residual forces on the ions was less than $1$~mRy$/$a.u.

\section{Spin states, breathing-mode distortions, and orbital order in unstrained bulk LCO}

In the LS ground state of bulk LCO, the $t_{2g}$ states are fully occupied 
and separated from the $e_g$ states by a band gap of $\sim 0.7$~eV (Fig.~\ref{fig:LCO-Bulk}; $0.45$~eV in LDA~\cite{RondinelliSpaldin:09}).
This value is in good agreement with experimental results ($\sim0.6$--$0.9$~eV~\cite{Abbate:93, Chainani:92}).
We find an O-Co-O distance of $3.88~\AA$ ($3.86~\AA$ from x-ray diffraction~\cite{Maris-LCO-OrbitalOrder:03}),
an octahedral volume of \smash{$9.8~\AA^3$},
and a bond angle of $160^\circ$ ($166^\circ$ in LDA~\cite{RondinelliSpaldin:09}).

The IS excited state ($82$~meV/Co higher in total energy than LS) is obtained
by transferring one electron from the minority spin $t_{2g}$ states to the majority spin $e_g$ states,
resulting in a local magnetic moment of $1.96~\muB$ ($1.8~\muB$ in LDA~\cite{RondinelliSpaldin:09})
at the equivalent Co sites.
The Fermi energy is located within the minority spin band gap, i.e., the system is half metallic.
This configuration with a single electron in the majority spin $e_g$ states resembles FM-polarized LaNiO$_3$
(being formally $d^7$).~\cite{ABR:11}
Interestingly, we find it to be unstable:
LCO in the IS state undergoes a transition to a phase with checkerboard charge disproportionation (CD) at the Co sites,
formally denoted as Co$^{3+}$ $\to$ Co$^{3+\delta}$ $+$ Co$^{3-\delta}$,
which lowers the total energy by $17$~meV/Co.
This is expressed in the variation of the local Co magnetic moments, $1.7/2.2~\muB$,
and accompanied by a structural breathing mode distortion (O-Co-O distances: $3.83/3.93~\AA$, octahedral volumes: \smash{$9.3/10.1~\AA^3$}).
A similar type of electronic reconstruction
is well known for nickelate films and heterostructures.~\cite{ABR:11, RENickelateReview:16, WrobelGeisler:18, GeislerPentcheva-LNOLAO:18}
For LCO, spin disproportionation has been suggested in model studies~\cite{Kunes-LCO-Disprop-Model:11}.

The CD causes a splitting of the $e_g$ manifold into two subsets [Fig.~\ref{fig:LCO-Bulk-OO}(a)],
similar as in (LaNiO$_3$)$_1$/(LaAlO$_3$)$_1(001)$ SLs~\citep{ABR:11, GeislerPentcheva-LNOLAO:18}.
In the majority spin channel this occurs at the Fermi energy,
and an indirect band gap emerges between the filled and the empty subset,
clearly visible in the band structure shown in Fig.~\ref{fig:LCO-Bulk}(c).
In the minority spin channel,
the two subsets range from $1$ to $2.2$~eV and from $2.2$ to $4$~eV, respectively [Fig.~\ref{fig:LCO-Bulk-OO}(a)].
The lower (upper) subset stems predominantly from the Co1 (Co2) sites; this sequence is reversed in the majority spin channel.
Interestingly, contributions of the Co $d_{x^2-y^2}$ and $d_{z^2}$ orbitals are similar in size in each subset,
which is indicative of the absence of Jahn-Teller effects.
No orbital order is observed among the $e_g$ states [Fig.~\ref{fig:LCO-Bulk-OO}(b)].

At the same time, sharp resonances appear in the minority spin DOS at $\sim 0.8$ and $1.55$~eV with distinct $t_{2g}$ orbital character: predominantly Co1 $d_{xy}+d_{yz}$ and Co2 $d_{xy}+d_{xz}$, respectively [Figs.~\ref{fig:LCO-Bulk}(b), \ref{fig:LCO-Bulk-OO}(a), \ref{fig:LCO-Bulk-OO}(b)]; each containing one hole.
This points to the simultaneous emergence of $t_{2g}$ $G$-type orbital order
that matches the CD checkerboard arrangement.

The present IS+CD+OO phase,
which shows orbital order exclusively for the hole in the minority spin $t_{2g}$ states
and accommodates the electron in the majority spin $e_g$ states via a charge and bond disproportionation mechanism,
is at variance with earlier suggested orbital order in both the $e_g$ and $t_{2g}$ states~\cite{Korotin:96}.
Notably, we find only negligible difference ($< 0.01~\AA$) in the three O-Co-O distances of each individual octahedron, i.e., almost perfectly symmetric octahedra.
In contrast, the impact of CD on the O-Co-O distances ($\sim 0.1~\AA$) is at least one order of magnitude larger.
Therefore, we suggest CD to be the reason for the experimentally observed distinct O-Co-O distances
($3.86~\AA$ at $90$~K splitting up symmetrically into $3.76/3.98~\AA$ at $295$~K~\cite{Maris-LCO-OrbitalOrder:03}, with some domains remaining NM, as it has been observed very recently for LCO films~\cite{Feng-LCO:19}),
instead of $e_g$ orbital order assisted by Jahn-Teller distortions.
Shimizu \textit{et al.}\ reported a symmetry preservation across the spin transition, which is incompatible with Jahn-Teller distortions~\cite{Shimizu:17}.
A previously suggested HS/LS mixed state with checkerboard order~\cite{Knizek-Ku-LCO:09, SeoDemkov:12, HsuBlahaWentzcovitch:12, Sterbinsky:18} exhibits the same space group as the IS+CD+OO phase,
but is substantially higher in energy than even the conventional IS state (without site disproportionation) if a larger fraction of Co ions gets thermally excited~\cite{Knizek-Ku-LCO:09}.
The similarity of experimentally observed and our predicted O-Co-O distances
provides additional evidence that the IS+CD+OO state
plays a role in the thermally driven spin transition of LCO.
The band gap arising due to site disproportionation offers a yet unexplored route to reconcile the otherwise metallic IS state with the experimentally observed low conductivity.~\cite{YanZhouGoodenough:04, Klie:07, Doi:14}
This holds in particular if the emerging FM domains are embedded in an insulating NM matrix~\cite{Feng-LCO:19}.
%
Finally, we note that a similar electronic structure is obtained for all systems in the IS state throughout this work.

\begin{figure}[t]
	\centering
	\includegraphics[]{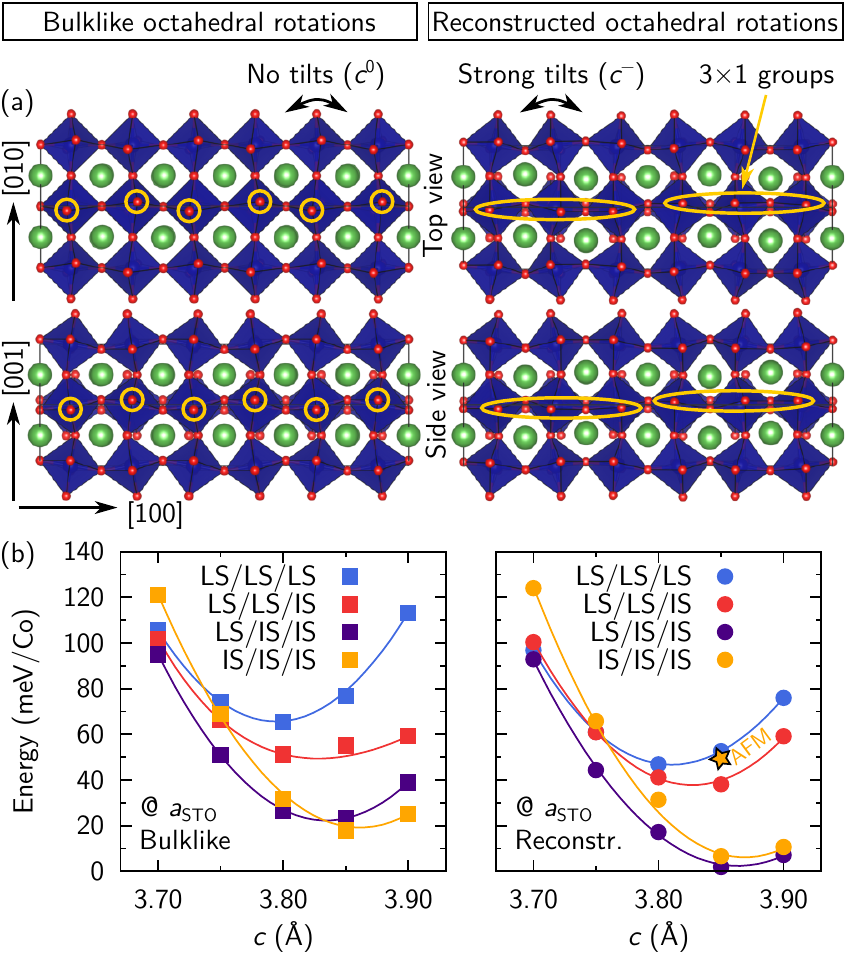}
	\caption{Stoichiometric model of LCO films on STO$(001)$. (a)~Top and side views representing schematically the $3 \times 1$ reconstruction of the octahedral rotation pattern. (b)~Total energy curves for bulklike and reconstructed octahedral rotations and four different superimposed ferromagnetic configurations as functions of the cell height~$c$. The star indicates the total energy of the IS/IS/IS system with antiferromagnetic order for comparison. The reconstructed octahedral rotations are more stable, irrespective of the magnetic order in the film, and ferromagnetism emerges purely due to tensile strain.}
	\label{fig:LCO-STO-Stoich}
\end{figure}

\section{\boldmath Strained LCO films on STO$(001)$}


Motivated by reports of a striped TEM pattern along the $[100]$ direction
appearing in LCO films grown on STO$(001)$,~\cite{Choi:12, Kwon:14, BiskupVarela:14, Mehta:15}
which is related to a short$/$short$/$long modulation of the La-La distances
($3.61/3.61/4.54~\AA$ instead of $3 \times 3.905~\AA$),~\cite{BiskupVarela:14}
we now discuss different types of $3 \times 1$ reconstructions
emerging in epitaxial LCO films on STO$(001)$
subject to $+2~\%$ tensile epitaxial strain.
Particularly, we explore structures with different oxygen concentrations.
In experiment, the films are insulating~\cite{Freeland-LCOSTO:08}
and exhibit FM order.~\cite{Fuchs:08, Freeland-LCOSTO:08, Mehta:09, Choi:12, BiskupVarela:14, Mehta:15, Qiao-LCO:15, Feng-LCO:19}

\subsection{Reconstructed octahedral rotations in stoichiometric films}

The first peculiar observation 
is a purely structural $3 \times 1$ reconstruction
of the rotational pattern of the CoO$_6$ octahedra
that is induced by tensile strain [Fig.~\ref{fig:LCO-STO-Stoich}(a)].
The bulklike $a^-b^-c^-$ octahedral rotations (actually $a^-b^-c^0$ due to strain)
turn out to be metastable and
are replaced by a pattern
in which the octahedra form groups of three with $a^+$ rotations.
Adjacent groups exhibit antiferrodistortive rotations around the $a \sim [100]$ axis.
Simultaneously, considerable octahedral rotations emerge
around the $c \sim [001]$ axis despite tensile strain,
resulting in an $a^+b^-c^-$ pattern inside each group.
This reconstruction lowers the total energy by $\sim 20$~meV/Co,
irrespective of the magnetic order in the LCO film,
as can be inferred from comparing the left and right panel in Fig.~\ref{fig:LCO-STO-Stoich}(b).
Since the reconstruction occurs for all considered magnetic orderings,
its origin is predominantly of ionic/electrostatic character and accommodates strain;
electronic effects are not the critical driving force.
Hence, stoichiometric LCO films strained on STO$(001)$ constitute an interesting example
for the complexity of emerging octahedral rotation patterns in transition metal oxides
and underline the importance of octahedral rotations in this system.

\begin{figure*}[t]
	\centering
	\includegraphics[]{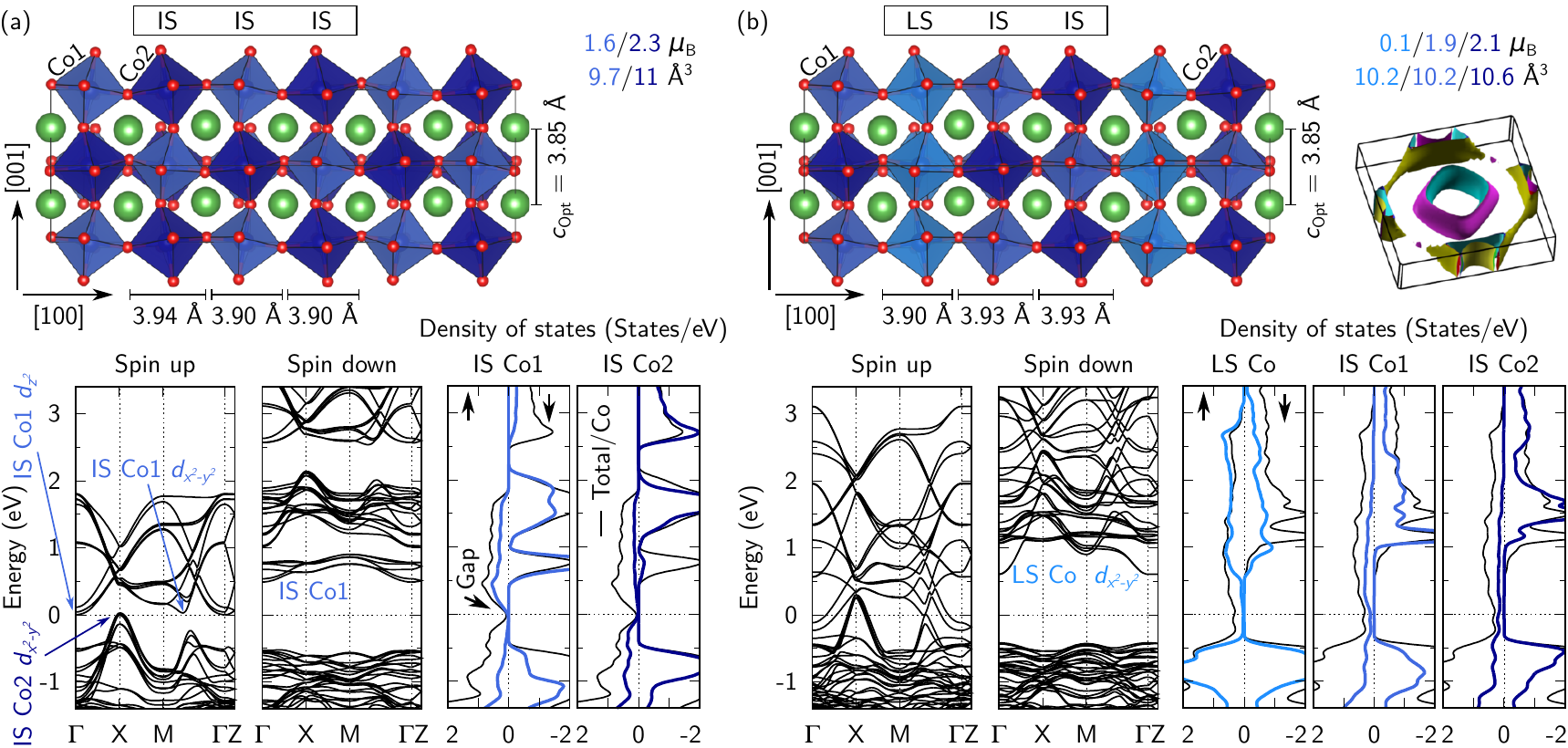}
	\caption{Magnetic modulations emerging in stoichiometric LCO films on STO$(001)$ in addition to the octahedral reconstruction (cf.~Fig.~\ref{fig:LCO-STO-Stoich}). In each case, the optimized atomic structure with different La-La distances, the spin-resolved band structure, the normalized total DOS per Co atom (black lines), and PDOS at the different Co sites (colored lines) are shown. (a)~The IS/IS/IS configuration exhibits an indirect band gap delimited by states of distinct IS Co character that is opened by site disproportionation. (b)~The LS/IS/IS configuration is semimetallic due to the overlap of IS Co $e_g$ states. Its Fermi surface has a distorted cylindrical/toroidal shape.}
	\label{fig:LCO-STO-Magnetism}
\end{figure*}

\subsection{Modulation of spin and orbital order in stoichiometric films}

Starting from the reconstructed octahedral rotation pattern,
we now explore the additional effect of different magnetic configurations
in LCO films on STO$(001)$.
We further optimize the ionic positions in each case.

\subsubsection{Emergence and modulation of ferromagnetic order}

The total energy curves in Fig.~\ref{fig:LCO-STO-Stoich}(b) show
that, starting from the LS/LS/LS state (i.e., absence of magnetism) with the highest energy,
by increasing the number of IS Co $(100)$ planes
the system gets continuously stabilized,
reaching its minimal energy for LS/IS/IS magnetic order.
Proceeding further to IS/IS/IS order destabilizes the system slightly.
Thus, the application of tensile strain is sufficient to induce FM order in LCO films,
and particularly the presence of oxygen vacancies is not required.
Moreover, the magnetic order is $3 \times 1$ modulated,
superimposing the octahedral reconstruction.
It is also noteworthy
that the total energy difference between LS/LS/IS and LS/IS/IS ($\sim 35$~meV/Co)
is much higher than between LS/LS/LS and LS/LS/IS ($\sim 9$~meV/Co),
i.e., there is no simple linear scaling with the number of IS Co $(100)$ planes.
Exemplarily for the IS/IS/IS system, antiferromagnetic (AFM) order was checked and found to be $43$~meV/Co higher in energy than FM order.
The relative stability of different magnetic states as a function of $U_\text{Co}$ is provided in the Supplemental Material.

\subsubsection{La-La distances, O-Co-O distances, and bond angles}

From Fig.~\ref{fig:LCO-STO-Stoich}(b) one can infer an optimized cell height of $c_\text{opt} \approx 3.85~\AA$.
This constitutes a considerable deviation from the volume conservation expectation,
which corresponds to $c \approx 3.68~\AA$.
While this may provide evidence for a negative Poisson's ratio of LCO,
experiments report a slightly lower value of $c \sim 3.8~\AA$~\cite{Mehta:15, Feng-LCO:19}, i.e., closer to our NM LS/LS/LS result [Fig.~\ref{fig:LCO-STO-Stoich}(b)].
This can be explained by (i)~the coexistence of FM and NM domains in experiment~\cite{Feng-LCO:19} and (ii)~a small overestimation of the lattice properties as typical for PBE$+U$~\cite{VermaGeislerPentcheva:19}.

Since the total energies of LS/IS/IS and IS/IS/IS magnetic order are very close
($\Delta E \sim 5$~meV/Co),
we compare both systems in Fig.~\ref{fig:LCO-STO-Magnetism}.
Variations of the La-La distances in the $[100]$ direction occur for both phases,
but amount to only $0.03$-$0.04~\AA$,
which is much smaller than the experimentally measured values.~\cite{BiskupVarela:14}
For IS/IS/IS order, they are caused by the reconstructed octahedral rotation pattern
and reproduce the short$/$short$/$long experimental trend.~\cite{Choi:12, Kwon:14, BiskupVarela:14}
In contrast, LS/IS/IS order
exhibits a short$/$long$/$long modulation of the La-La distances,
which is at qualitative variance with experiment.
LS/LS/IS order qualitatively reproduces the experimental trend,
but is too high in energy [Fig.~\ref{fig:LCO-STO-Stoich}(b)].
We will see in the following that
the stoichiometric phase stabilizes only at high oxygen pressure,
while the experiments are usually performed at low oxygen pressure.

\begin{table}[b]
	\centering
	\vspace{-1.5ex}
	\caption{\label{tab:StructData}Overview of site-resolved O-Co-O distances along different directions (cf.~Fig.~\ref{fig:HubbardU-Geometry}), CoO$_6$ octahedral volumes $V$, and local Co magnetic moments $m$ of different LCO systems on STO$(001)$ and of bulk LCO for reference. Note that the different reconstruction mechanisms discussed in the text decouple all three spatial directions.}
	\begin{ruledtabular}
	\begin{tabular}{lccccc}
		Site & \multicolumn{3}{c}{O-Co-O distance ($\AA$)} & $V$ ($\AA^3$) & $m$ ($\muB$)	\\
		& $[100]$ & $[010]$ & $[001]$ & &	\\
		\hline
		\multicolumn{6}{l}{LaCoO$_{3}$ (strained), IS/IS/IS magnetic order (Fig.~\ref{fig:LCO-STO-Magnetism})}	\\
		IS Co1	& $3.89$	& $3.89$ & $3.85$ & $9.7$	& $1.6$	\\
		IS Co2	& $4.06$	& $4.06$ & $4.00$ & $11$	& $2.3$	\\
		\hline
		\multicolumn{6}{l}{LaCoO$_{3}$ (strained), LS/IS/IS magnetic order (Fig.~\ref{fig:LCO-STO-Magnetism})}	\\
		LS Co	& $3.89$	& $3.99$ & $3.95$ & $10.2$ & $0.1$	\\
		IS Co1	& $4.00$	& $3.93$ & $3.87$ & $10.2$ & $1.9$	\\
		IS Co2	& $4.03$	& $3.99$ & $3.96$ & $10.6$ & $2.1$	\\
		\hline
		\multicolumn{6}{l}{LaCoO$_{2.67}$ (strained), ordered oxygen vacancies (Fig.~\ref{fig:LCO-STO-OxVac})}	\\
		LS Co oct.	& $3.99$	& $3.87$ & $3.89$ & $10$ & $0.3$	\\
		HS Co oct.	& $4.30$	& $4.06$ & $4.12$ & $12$ & $2.6$	\\
		\hline
		\multicolumn{6}{l}{LaCoO$_{2.5}$ (strained), brownmillerite (Fig.~\ref{fig:LCO-STO-BM})}	\\
		HS Co oct.	& $4.52$	& $3.95$ & $4.11$ & $12.2$ & $\pm 2.5$	\\
		\hline
		\multicolumn{6}{l}{LaCoO$_{3}$ bulk, LS state (Fig.~\ref{fig:LCO-Bulk})}	\\
		LS Co	& & $3.88$ & & $9.8$ & $0$	\\
		\hline
		\multicolumn{6}{l}{LaCoO$_{3}$ bulk, IS+CD+OO state (Fig.~\ref{fig:LCO-Bulk})}	\\
		IS Co1	& & $3.83$ & & $9.3$	& $1.7$	\\
		IS Co2	& & $3.93$ & & $10.1$	& $2.2$	\\
	\end{tabular}
	\end{ruledtabular}
\end{table}

In an earlier DFT study, Kwon \textit{et al.}\
reported a strain-induced magnetic ground state with LS/LS/HS order
and La-La distances ($3.83/3.83/4.17~\AA$) in line with the experimental trend~\cite{Kwon:14}.
In our large supercells and with explicit treatment of octahedral rotations,
this phase could only be obtained under application of additional constraints to the total magnetization.
Once these constraints were lifted, the HS Co ions relaxed to an IS state,
which implies that the LS/LS/HS phase is not even metastable.
The different observations in previous work probably stem from the use of small supercells and the neglect of octahedral rotations,
which are known to strongly impact magnetism in LCO~\cite{RondinelliSpaldin:09}.
%
Moreover, we explored the mixed HS/LS phase with checkerboard order~\cite{Knizek-Ku-LCO:09, SeoDemkov:12, HsuBlahaWentzcovitch:12, Sterbinsky:18} (see Supplemental Material).
Optimization rendered $c_\text{opt} \approx 3.85~\AA$.
We found this phase to be considerably lower in energy than expected from bulk extrapolation~\cite{RondinelliSpaldin:09}, which is indicative of a strong stabilizing effect.
This is in line with earlier PBE$+U$ results for bulk LCO~\cite{Knizek-Ku-LCO:09}.
However, the IS/IS/IS and LS/IS/IS phases, both containing IS+CD+OO, are substantially lower in energy ($\sim 60$~meV/Co). We verified this for a large range of $U_\text{Co}$ values and different DFT$+U$ techniques~\cite{QE-LDA-U:05, LiechtensteinAnisimov:95} (see Supplemental Material).

\begin{figure}[t]
	\centering
	\includegraphics[]{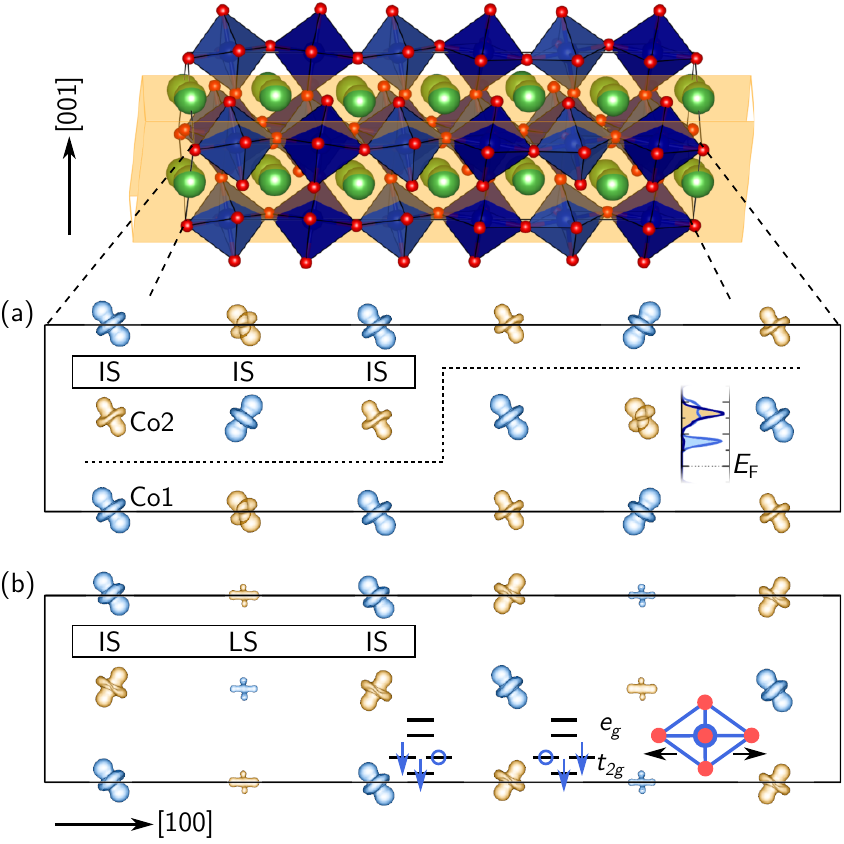}
	\caption{Complex orbital order of the $t_{2g}$ holes in LCO films on STO$(001)$ (cf.~Fig.~\ref{fig:LCO-STO-Magnetism}), visualized by superimposed ILDOS isosurfaces (blue: integrated Co1 peak; orange: integrated Co2 peak; cf.~Fig.~\ref{fig:LCO-Bulk-OO}). (a)~For IS/IS/IS, instead of bulklike $G$-type orbital order, the reconstructed octahedral rotation pattern causes a shift in the orbital order pattern that occurs every third Co $(100)$ plane (indicated by the dashed lines). Three different features can be identified. (b)~For LS/IS/IS, the IS Co ions show perfect $G$-type orbital order, interrupted only by the LS Co $(100)$ planes. The energy diagrams indicate the strain-induced splitting and alternating occupation of the IS Co~$3d$ minority spin orbitals, the empty circle denoting the hole. The figure has been optimized for clarity; see Supplemental Material for the original data.}
	\label{fig:LCO-STO-OrbitalOrder}
\end{figure}

Table~\ref{tab:StructData} lists the octahedral O-Co-O distances in the $[100]$, $[010]$, and $[001]$ directions.
For IS/IS/IS, the O-Co-O distances in the $(001)$ plane are expanded to $3.89$ and \smash{$4.06~\AA^3$}, varying around the STO substrate lattice constant \smash{$3.905~\AA^3$}
and reflecting the bond disproportionation (similar to our observations for bulk LCO).
The resulting octahedral volumes are $9.7$ and \smash{$11~\AA^3$} for IS Co1 and IS Co2, respectively.
For LS/IS/IS, the situation is more complex:
The octahedra show different O-Co-O distances along all three directions owing to the octahedral connectivity
and the LS Co $(100)$ layers,
resulting in octahedral volumes of $10.2$, $10.2$, and \smash{$10.6~\AA^3$} for LS Co, IS Co1, and IS Co2, respectively.
The bond disproportionation is reduced with respect to the IS/IS/IS case,
which is also reflected in the smaller local magnetic moment difference between distinct IS Co sites.

Without the octahedral reconstruction,
the Co-O-Co bond angles
of the IS/IS/IS configuration
are $161^\circ$-$162^\circ$ in the $(001)$ plane, which is similar to bulk LCO,
and $155^\circ$-$157^\circ$ in the perpendicular $[001]$ growth direction,
i.e., the octahedral rotations are significantly larger than in bulk LCO.
Nevertheless, we observe little to no octahedral rotations around the $[001]$ axis.
With the octahedral reconstruction,
the Co-O-Co bond angles
of the IS/IS/IS configuration
are $157^\circ$-$159^\circ$ in the $(001)$ plane
as well as $156^\circ$-$158^\circ$ in the $[001]$ direction.
Particularly, octahedral rotations around the $[001]$ axis appear despite tensile strain.
Additional magnetic modulation in the LS/IS/IS case leads to larger spread of the Co-O-Co bond angles,
namely $156^\circ$-$161^\circ$ in the $(001)$ plane
and $155^\circ$-$160^\circ$ in the $[001]$ growth direction.

\subsubsection{Electronic properties and orbital order}

The influence of epitaxial strain on the electronic properties of LCO films on STO$(001)$ is displayed in Fig.~\ref{fig:LCO-STO-Magnetism}.
We find the IS/IS/IS phase to be semiconducting due to site disproportionation,
similar to bulk LCO in the IS+CD+OO phase reported above.
An indirect band gap separates majority-spin IS Co1 states acting as conduction band minimum
(CBM, having $d_{z^2}$ character at $\Gamma$ and $d_{x^2-y^2}$ character between $M$ and $\Gamma$)
and majority-spin IS Co2 states acting as valence band maximum (VBM, having $d_{x^2-y^2}$ character at $X$)
[Fig.~\ref{fig:LCO-STO-Magnetism}(a)].
In contrast, we observe a semimetallic phase for LS/IS/IS
due to an overlap of the electron pocket at $\Gamma$
and the hole pocket around $X$ [Fig.~\ref{fig:LCO-STO-Magnetism}(b)].
It is peculiar that the periodic appearance of insulating LS Co $(100)$ planes
(i)~leads to a slightly lower total energy
(ii)~despite inducing semimetallicity in the system
(iii)~by causing IS Co $e_g$ states to overlap that otherwise would be separated.
However, the conductivity is still impeded by the insulating LS planes.
From the structural data and the similarity of the Co magnetic moments ($1.9$ and $2.1~\muB$, Table~\ref{tab:StructData})
we conclude that the CD is inhibited by the LS Co $(100)$ planes,
thereby preventing a band gap to open.
The Fermi surface shows a distorted cylindrical/toroidal shape and thus a reconstructed topology compared to bulk LCO in the conventional metallic IS phase [cf.~Fig.~\ref{fig:LCO-Bulk}(b)].

By comparing the PDOS of the IS/IS/IS and LS/IS/IS configurations shown in Fig.~\ref{fig:LCO-STO-Magnetism}
one can see that the IS Co1 peak in the minority spin channel
shifts from $0.8$ to $1.1$~eV,
a signature of quantum confinement caused by the insulating LS Co $(100)$ planes.
In contrast, the IS Co2 peak remains largely $\sim 1.6$~eV above the Fermi energy.
These peaks can be related to bulk LCO in the IS+CD+OO phase [cf.~Figs.~\ref{fig:LCO-Bulk}(b) and~\ref{fig:LCO-Bulk-OO}(a)].

As one can infer from Fig.~\ref{fig:LCO-STO-OrbitalOrder},
we find orbital order at the IS Co sites in LCO films on STO$(001)$
that is considerably impacted by
the $3 \times 1$ reconstructions
of the octahedral rotation pattern
and the magnetic order.
A fundamental difference to bulk LCO is that
tensile strain and the resulting basal expansion of the CoO$_6$ octahedra
lower the energy of one of the minority spin $t_{2g}$ orbitals, which is therefore always occupied,
whereas the hole alternately occupies one of the remaining two minority spin $t_{2g}$ orbitals
(see energy diagrams in Fig.~\ref{fig:LCO-STO-OrbitalOrder}).
For LS/IS/IS, the IS Co ions show perfect $G$-type orbital order,
which is interrupted only by the LS Co $(100)$ planes.
For IS/IS/IS, the orbital order does not match the CD pattern as in bulk LCO.
Instead, the reconstructed octahedral rotations cause a shift in the orbital order pattern
along the $[001]$ direction
that occurs every third Co $(100)$ plane,
leading to a more complex orbital order.

\subsection{\boldmath Ordered oxygen vacancies in LaCoO$_{2.67}$ on STO$(001)$}

\begin{figure}[t]
	\centering
	\includegraphics[]{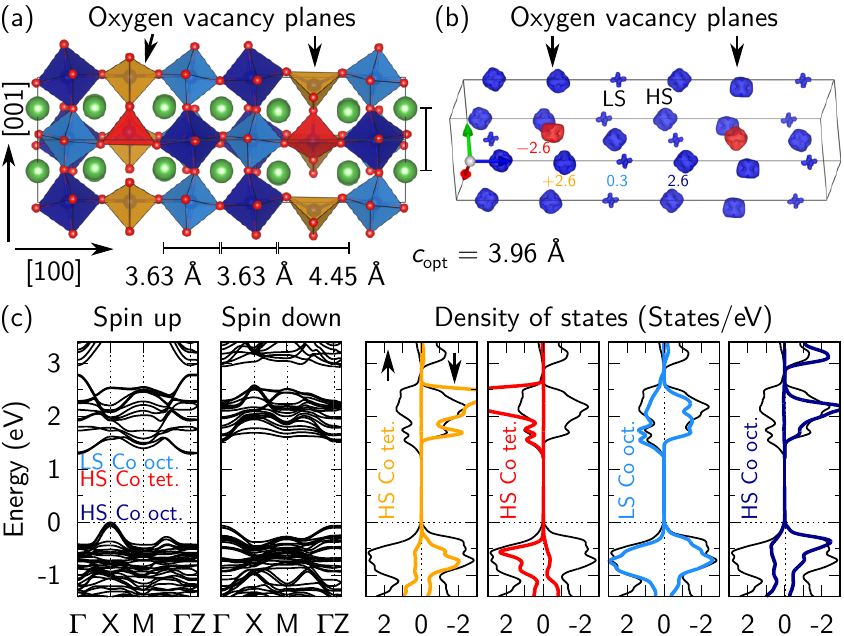}
	\caption{(a)~Ordered oxygen vacancy model of LCO/STO$(001)$ films, formally LaCoO$_{2.67}$. CoO$_6$ octahedra (CoO$_4$ tetrahedra) are depicted in light/dark blue (orange/red). The La-La distances and the optimized cell height are provided. (b)~The spin density visualizes the AFM order in the tetrahedron planes (i.e., oxygen vacancy planes) and the checkerboard LS/HS Co$^{3+}$/Co$^{2+}$ charge order with FM spin alignment in the octahedron planes. The numbers represent Co magnetic moments ($\muB$). (c)~In addition, the spin-resolved band structure, the normalized total DOS per Co atom (black lines), and PDOS at the different Co sites (colored lines) are shown.}
	\label{fig:LCO-STO-OxVac}
\end{figure}

We next turn to the reduced systems with oxygen vacancies.
Figure~\ref{fig:LCO-STO-OxVac}(a) shows a model~\cite{BiskupVarela:14} of LCO films epitaxially grown on STO$(001)$
with ordered oxygen vacancies in every third Co $(100)$ plane,
formally LaCoO$_{2.67}$.
Our optimization resulted in $c_\text{opt} = 3.96~\AA$, which is larger than in the stoichiometric case and related to the partial Co reduction. 
We confirmed this expansion by additional simulations employing the Vienna Ab initio simulation package~\cite{PAW:94,kresse1996b,USPP-PAW:99} (see Supplemental Material),
but note that the concentration and geometry of the oxygen vacancies~\cite{FumegaPardo:17}
as well as the choice of the exchange-correlation functional
may impact the cell height.
The planar La-La distances
are contracted around the CoO$_6$ octahedra and expanded around the CoO$_4$ tetrahedra
[i.e., the oxygen vacancy $(100)$ planes]
and amount to $3.63/3.63/4.45~\AA$,
which are close to the measured values $3.61/3.61/4.54~\AA$.~\cite{BiskupVarela:14}
The tetrahedra contain HS Co$^{2+}$ ($S=3/2$) with local magnetic moments of $\pm 2.6~\muB$, i.e., AFM order emerges in the tetrahedron planes [Fig.~\ref{fig:LCO-STO-OxVac}(b)].
In contrast, the octahedra exhibit a checkerboard LS Co$^{3+}$ ($S=0$) / HS Co$^{2+}$ ($S=3/2$) charge order with FM spin alignment, the local magnetic moments being $0.3$ and $2.6~\muB$, respectively.
The total magnetic moment of the supercell is $12~\muB$.
We observe a concomitant bond disproportionation,
resulting in octahedral volumes of $\sim 10$ and \smash{$12~\AA^3$} for LS and HS Co, respectively.
The volume of the tetrahedra is always \smash{$\sim 3.93~\AA^3$}.
The octahedral O-Co-O distances listed in Table~\ref{tab:StructData}
reveal a strong elongation along the $[100]$ direction ($3.99/4.30~\AA$). 
The tetrahedral Co-O bond lengths are
$1.95$-$1.99~\AA$ along the $[100]$ direction and
$2.01$-$2.03~\AA$ in the $(100)$ plane.

This system shows a Peierls instability,
driving a modulation of the La-La distances~\cite{BiskupVarela:14}.
In conjunction with the electronic reconstruction of the Co valence state, it accommodates the electrons released by the introduced oxygen vacancies, resulting in an insulating state [Fig.~\ref{fig:LCO-STO-OxVac}(c)].
We find the band edges to be located in the majority spin channel,
and an indirect $X \rightarrow \Gamma$ band gap of $1.29$~eV,
which is almost twice as large as in LCO bulk [cf.~Fig.~\ref{fig:LCO-Bulk}(c)].
The band gap in the minority spin channel is even larger.
The VBM is dominated by states with octahedral HS Co character,
showing a similar characteristic feature at the $X$~point as in the stoichiometric case (cf.~Fig.~\ref{fig:LCO-STO-Magnetism}),
whereas the CBM is governed by states with octahedral LS Co and some tetrahedral HS Co character.

\begin{figure}[t]
	\centering
	\includegraphics[]{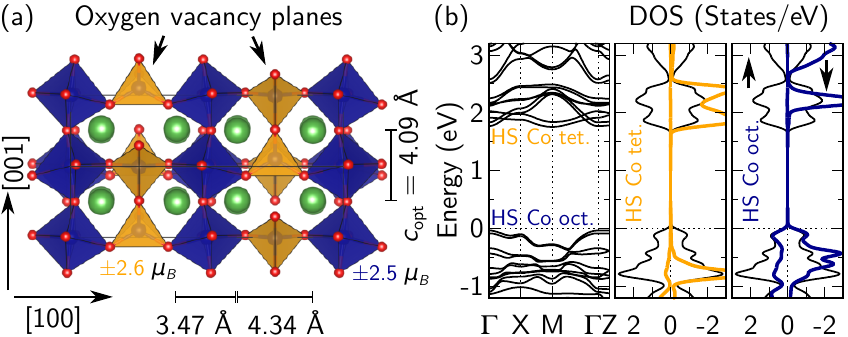}
	\caption{(a)~Brownmillerite LaCoO$_{2.5}$ films on STO$(001)$. CoO$_6$ octahedra (CoO$_4$ tetrahedra) are depicted in blue (orange). The La-La distances, the optimized cell height, and the Co magnetic moments are given. (b)~Normalized total DOS per Co atom (black lines) and projections on Co sites with positive magnetic moment (colored lines; mind the $G$-type AFM order). Moreover, the band structure is provided.}
	\label{fig:LCO-STO-BM}
\end{figure}

\subsection{\boldmath Brownmillerite LaCoO$_{2.5}$ on STO$(001)$}

The discussed LaCoO$_{2.67}$ model is actually an intermediate structure
between the perovskite LaCoO$_3$
and the brownmillerite LaCoO$_{2.5}$~\cite{Hansteen-LCO-BM:98},
which exhibits ordered oxygen vacancies in every second Co $(100)$ plane [Fig.~\ref{fig:LCO-STO-BM}(a)].
Although its $2 \times 1$ pattern does not fit the $3 \times 1$ experimental observation,
it is worthwhile to inspect this system for completeness.
According to our calculations, 
epitaxial LaCoO$_{2.5}$ films on STO$(001)$ are $G$-type AFM ordered
with local Co magnetic moments of $\pm 2.6$ (tetrahedra) and $\pm 2.5~\muB$ (octahedra)
and a total energy difference to FM order of $86$~meV/Co.
Both tetrahedral and octahedral Co are HS Co$^{2+}$
[note the similarities in the PDOS comparing Figs.~\ref{fig:LCO-STO-OxVac}(c) and~\ref{fig:LCO-STO-BM}(b)].
Thus, the differences to the LaCoO$_{2.67}$ system
are the absence of LS Co$^{3+}$ ions
and AFM order among the octahedral Co ions instead of FM order.
As a consequence, we find the epitaxial films to be insulating as well.
The band gap amounts to $1.75$~eV,
which is significantly larger than for the LaCoO$_{2.67}$ structure [cf.~Fig.~\ref{fig:LCO-STO-OxVac}(c)].
The VBM is dominated by states with octahedral Co character,
whereas the CBM is governed by states with tetrahedral Co character.
The indirect $\Gamma \rightarrow M$ band gap is only slightly smaller than the direct $\Gamma \rightarrow \Gamma$ band gap.
The optimized cell height is $c_\text{opt}=4.09~\AA$,
which is even further expanded with respect to the stoichiometric and the LaCoO$_{2.67}$ cases.
This can be understood from the fact that LaCoO$_{2.5}$
is under \textit{compressive} strain on STO$(001)$~\cite{Hansteen-LCO-BM:98}.
The two distinct La-La distances in the $[100]$ direction are $3.47$ and $4.34~\AA$,
both being smaller than in the LaCoO$_{2.67}$ case.
The octahedral volume is \smash{$12.2~\AA^3$}, whereas the volume of the tetrahedra is \smash{$3.98~\AA^3$}.
Table~\ref{tab:StructData} lists the octahedral O-Co-O distances in the $[100]$, $[010]$, and $[001]$ directions,
revealing even stronger octahedral distortions than observed in the LaCoO$_{2.67}$ case.
The Co-O bond lengths in the tetrahedra are
$1.93$-$1.94~\AA$ along the $[100]$ direction and
$2.02$-$2.06~\AA$ in the $(100)$ plane.

\subsection{Phase diagram as function of the oxygen pressure}

\begin{figure}[t]
	\centering
	\includegraphics[]{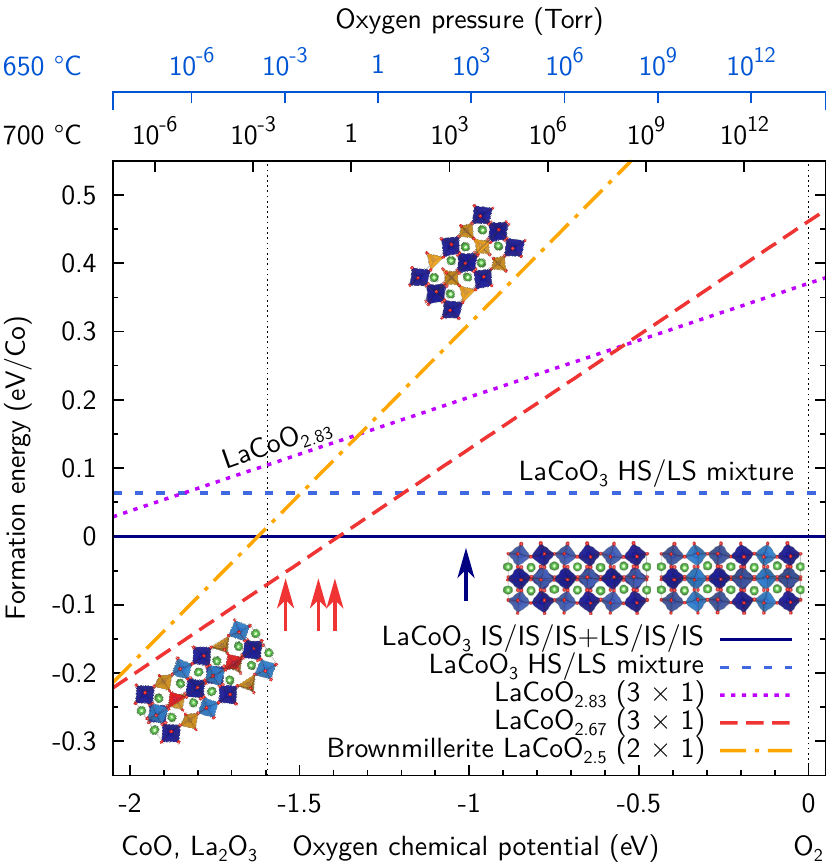}
	\caption{Phase diagram for different LCO systems strained on STO$(001)$. The colored lines represent the corresponding film formation energies $E_\text{f}(\mu_\text{O})$ (lower means more stable). The horizontal axes display the oxygen chemical potential $\mu_\text{O}$ and the related oxygen pressure at two different typical growth temperatures. The left (right) vertical dashed line depicts oxygen-poor (oxygen-rich) growth conditions. The arrows mark typical growth conditions used for LCO films on STO$(001)$.~\cite{Fuchs:08, Mehta:09, Choi:12, Kwon:14, BiskupVarela:14, Mehta:15} }
	\label{fig:Thermodynamics}
\end{figure}

In the context of \textit{ab initio} thermodynamics,~\cite{AbInitioThermodynamics}
the film formation energies
as shown in Fig.~\ref{fig:Thermodynamics}
depend on variables characterizing the chemical environment in which growth takes place;
here, they only depend on the oxygen chemical potential: $E_\text{f}(\mu_\text{O})$.
Thermodynamic equilibrium is described by the condition $E_\text{LaCoO$_{3}$} = \mu_\text{La} + \mu_\text{Co} + 3 \mu_\text{O}$.
%
We set $E_\text{LaCoO$_{3}$} = E_\text{LaCoO$_{3}$}^\text{@ STO}$
to the total energy of the stoichiometric LCO film on STO$(001)$ with LS/IS/IS magnetic order and reconstructed octahedral rotation pattern [cf.~Fig.~\ref{fig:LCO-STO-Magnetism}(b)].
Hence, the formation energy of this film is used as reference
and corresponds to a horizontal line in the diagram.
The energy of the IS/IS/IS phase is very similar and thus coincides with this line.
The mixed HS/LS phase is $\sim 60$~meV/Co less stable (see Supplemental Material).
The LS/LS/HS phase has an even higher energy ($\sim 140$~meV/Co, not shown).
For the oxygen-deficient structures we derive:
\begin{align*}
  E_\text{f}(\mu_\text{O}) &= E_\text{LaCoO$_{3-\delta}$}^\text{@ STO} - E_\text{LaCoO$_{3}$}^{} + \delta \cdot \mu_\text{O} \text{.}
\end{align*}
If $\mu_\text{O}$ is reduced below a certain threshold,
LCO will start to decompose into monoclinic CoO~\cite{Schroen:12}
and hexagonal La$_2$O$_3$,~\cite{La2O3:98}
which defines the oxygen-poor limit.
In contrast, the oxygen-rich limit is given by the energy of an O$_2$ molecule.
Hence,
\begin{equation*}
  2 E_\text{LaCoO$_3$} - 2 E_\text{CoO} - E_\text{La$_2$O$_3$} < \mu_\text{O} < \frac{1}{2} E_\text{O$_2$}
\text{.}
\end{equation*}
Assuming that oxygen forms an ideal-gas-like reservoir during sample growth,
its chemical potential and pressure can be related by~\cite{ReuterScheffler:01, PentchevaScheffler:05, MulakaluriPentchevaScheffler:09}
\begin{equation*}
  \label{eq:ChemPotPressure}
  \mu_\text{O}(T, p) = \mu_\text{O}(T, p^{\circ}) + \frac{1}{2} k_\text{B} T \ln \left( \frac{p}{p^{\circ}} \right)
\text{.}
\end{equation*}
Here we use the values for $\mu_\text{O}(T, p^{\circ})$ as tabulated in Ref.~\onlinecite{ReuterScheffler:01} for standard pressure $p^{\circ} = 760$~Torr.

\begin{table}[b]
	\centering
	\vspace{-1.5ex}
	\caption{\label{tab:ExpGrowthConditions}Overview of experimental growth conditions (oxygen pressure~$p$ and temperature~$T$) and measured Curie temperatures $T_\text{C}$ of LCO/STO$(001)$ films reported in the literature, and whether or not a $3 \times 1$ reconstruction has been observed in subsequent TEM analysis (--- means ``not investigated'').}
	\begin{ruledtabular}
	\begin{tabular}{lccccc}
		Reference & $p$ (mTorr) & $T$ (${}^{\circ}$C) & $T_\text{C}$ (K) & $3 \times 1$	\\
		\hline
		Fuchs \textit{et al.}~\cite{Fuchs:08}				& $675 \times 10^3{}^\dagger$ 		& $650$; $500^\dagger$ & $\sim 85$	& --- \\
		Freeland \textit{et al.}~\cite{Freeland-LCOSTO:08}	& $1$; $750 \times 10^3{}^\dagger$	& $750$; $580^\dagger$ & $\sim 80$	& --- \\
		Mehta \textit{et al.}~\cite{Mehta:09}				& $10$, $320$ 		& $700$ & $\sim 80^\ddagger$	& --- \\
		Choi \textit{et al.}~\cite{Choi:12}					& $100$ 			& $700$ & $\sim 80$				& yes \\
		Kwon \textit{et al.}~\cite{Kwon:14}					& $100$ 			& $700$ & ---					& yes \\
		Bi\v{s}kup \textit{et al.}~\cite{BiskupVarela:14}	& $320$ 			& $700$ & $\sim 80$ 			& yes \\
		Mehta \textit{et al.}~\cite{Mehta:15}				& $320$		 		& $700$ & $\sim 76$-$85$	& yes \\
		Qiao \textit{et al.}~\cite{Qiao-LCO:15}				& $200$; $200 \times 10^3{}^\dagger$ 			& $650$ & $\sim 80$ 			& no \\
		Feng \textit{et al.}~\cite{Feng-LCO:19}				& $190$ 			& $750$ & $\sim 85^*$ 			& no \\
	\end{tabular}
	\end{ruledtabular}
	${}^\dagger$ Employed during post-growth annealing.	\\
	${}^\ddagger$ Transition to FM order only observed for $p=320$~mTorr. \\
	${}^*$ Maximally $50~\%$ of the LCO film exhibit FM order.
\end{table}

\begin{table}[b]
	\centering
	\vspace{-1.5ex}
	\caption{\label{tab:ModelComparison}Comparison of first-principles results for different phases of LCO films strained on STO$(001)$.}
	\begin{ruledtabular}
	\begin{tabular}{lccc}
		Model & Stability & $3 \times 1$ & Electronic properties\\
		\hline
		\multicolumn{4}{l}{Oxygen-vacancy structures}	\\
		LaCoO$_{2.67}$	& yes & yes~\cite{BiskupVarela:14} & FM, insulating~\cite{BiskupVarela:14}	\\
		LaCoO$_{2.83}$	& no & yes~\cite{FumegaPardo:17} & FM, insulating~\cite{FumegaPardo:17}	\\
		LaCoO$_{2.5}$	& no${}^*$ & no & AFM, insulating	\\
		\hline
		\multicolumn{4}{l}{Stoichiometric structures}	\\
		IS/IS/IS & yes & weak$^\ddagger$ & FM, semiconducting	\\
		LS/IS/IS & yes & weak & FM, semimetallic${}^\dagger$	\\
		HS/LS mixture & no & weak$^\ddagger$ & FM, insulating~\cite{SeoDemkov:12, HsuBlahaWentzcovitch:12, Sterbinsky:18}	\\
		LS/LS/HS & no & yes~\cite{Kwon:14} & FM, insulating~\cite{Kwon:14}	\\
	\end{tabular}
	\end{ruledtabular}
	${}^\dagger$ Insulating LS planes substantially reduce conductivity.	\\
	${}^\ddagger$ Owing to the reconstructed octahedral rotation pattern. \\
	${}^*$ Impeded by preferential formation of competing oxides. \\
\end{table}

Experimental growth of LCO films on STO$(001)$
is usually carried out at around $650$-$750$~${}^\circ$C
and $1$-$320$~mTorr oxygen pressure,
as summarized in Table~\ref{tab:ExpGrowthConditions}.
Fuchs \textit{et al.},~\cite{Fuchs:08}
Freeland \textit{et al.},~\cite{Freeland-LCOSTO:08}
and Qiao \textit{et al.}~\cite{Qiao-LCO:15}
exposed the samples to a $200$-$750$~Torr oxygen atmosphere after deposition;
nevertheless, they observed a similar Curie temperature as Mehta \textit{et al.},~\cite{Mehta:09, Mehta:15}
Choi \textit{et al.},~\cite{Choi:12}
Bi\v{s}kup \textit{et al.},~\cite{BiskupVarela:14}
and Feng \textit{et al.}~\cite{Feng-LCO:19}
The relatively high temperatures employed during the growth process
legitimate our thermodynamic approach,
which is strictly valid only in equilibrium.~\cite{Geisler:12, Suzuki:13}

The phase diagram in Fig.~\ref{fig:Thermodynamics} shows that for high oxygen pressure during growth (corresponding to $\mu_\text{O} > -1.4$~eV)
epitaxial LCO films in perovskite structure form on STO$(001)$
that undergo different kinds of structural, electronic, and magnetic $3 \times 1$ reconstructions as discussed above.
Since the energy difference between the IS/IS/IS and the LS/IS/IS phase amounts to only a few meV/Co and depends also on $U_\text{Co}$ (see Supplemental Material), external influences (e.g., dilute impurities such as oxygen vacancies, TEM sample preparation conditions, proximity effects of differently magnetized domains, or the polarity of the LCO/STO$(001)$ interface) can impact which phase stabilizes.
Moreover, the La-La distance modulation is only weak in both phases (Fig.~\ref{fig:LCO-STO-Magnetism}, Table~\ref{tab:ModelComparison}).
This may offer an explanation why the $3 \times 1$ reconstruction of LCO films strained on STO$(001)$ is sometimes \textit{not} observed~\cite{Qiao-LCO:15, Feng-LCO:19}.
Both phases exhibit a low conductivity, in particular the semiconducting IS/IS/IS phase, but also the LS/IS/IS phase due to the insulating LS planes.
Moreover, the coexistence of FM and insulating NM domains in LCO films may add to the resistivity of the samples~\cite{Feng-LCO:19}.

For lower oxygen pressure (corresponding to $-2.1~\text{eV} < \mu_\text{O} < -1.4$~eV),
as typically used in those experimental studies that report the $3 \times 1$ striped TEM pattern,~\cite{Choi:12, Kwon:14, BiskupVarela:14, Mehta:15}
film structures with ordered oxygen vacancies in every third Co $(100)$ plane (LaCoO$_{2.67}$) are the most stable (Fig.~\ref{fig:Thermodynamics}).
The growth parameters employed in these experiments are near our oxygen-poor limit,
but still well within the interval where LCO is stable (i.e., $\mu_\text{O} > -1.6$~eV).
We thus conclude
that the striped TEM pattern
originates most likely from $(100)$ planes of ordered oxygen vacancies.
This is also consistent with the measured insulating nature of LCO/STO$(001)$ films~\cite{Freeland-LCOSTO:08}.
%
We complete the picture by exploring the LaCoO$_{2.83}$ phase reported by Fumega and Pardo~\cite{FumegaPardo:17} (see Supplemental Material) and found it to be
less stable than LaCoO$_3$ or LaCoO$_{2.67}$, irrespective of the growth conditions (Fig.~\ref{fig:Thermodynamics}).
The phase diagram was compiled by using the most stable oxygen vacancy configurations according to Refs.~\onlinecite{BiskupVarela:14,FumegaPardo:17}. Further arrangements of oxygen vacancies as well as the impact of isolated and charged vacancies~\cite{Freysoldt-PointDefects:14, Janotti-TiO2:10, zhang:91} should be considered in future work.

It is interesting to compare these cases
to the epitaxial growth of LaCoO$_{2.5}$ films in brownmillerite structure,
i.e., a system with further increased oxygen vacancy density.
We find it to require a much lower oxygen pressure (corresponding to $\mu_\text{O} < -2.1$~eV)
that is already located in the regime where LCO is no longer stable and tends to decompose.
This is related to the relatively high formation energy of oxygen vacancies in La-based perovskites.~\cite{CurnanOxVac:14}
Hence, we expect epitaxial growth of LaCoO$_{2.5}$/STO$(001)$ to be difficult.
The situation is different in Sr-based cobaltates, which have readily been grown
in different stoichiometries on STO$(001)$.~\cite{SrCoO3-sponge-Jeen:13}

\vspace{-1.5ex}

\section{Summary}

By using density functional theory calculations
with a Coulomb repulsion term, we investigated
structural, electronic, and magnetic reconstruction mechanisms
as well as the impact of ordered oxygen vacancies and different octahedral rotation patterns
in epitaxial LaCoO$_{3}$ films grown on SrTiO$_3(001)$.
A moderate Hubbard-$U=3$~eV acting at the Co $3d$ states consistently provides the proper nonmagnetic and insulating ground state for bulk LaCoO$_3$.

For bulk LaCoO$_3$ in the intermediate-spin state
we reported a novel phase in which the excited electron is accommodated in the $e_{g}$ states via a charge- and bond-disproportionation mechanism,
bearing similarities to nickelate systems,
while orbital order occurs exclusively for the hole in the $t_{2g}$ manifold.
Breathing-mode distortions lead to O-Co-O distances that compare well with experimental x-ray diffraction measurements at elevated temperature.
In contrast, Jahn-Teller distortions were found to be suppressed.
The emergent band gap offers a so far unexplored route to reconcile the
otherwise metallic intermediate-spin state with the experimentally observed
low conductivity during the thermally driven spin transition of LaCoO$_3$.

For stoichiometric perovskite films on SrTiO$_3(001)$,
we found two novel competing ground-state candidates:
a $3 \times 1$ spin-reconstructed and semimetallic phase
with a peculiar reconstruction of the octahedral rotations,
charge and orbital order, and a Fermi surface topology distinct from bulk,
and a semiconducting phase of intermediate-spin magnetic order.
This shows that ferromagnetism emerges in epitaxial LaCoO$_3$ films even without oxygen vacancies,
purely by application of tensile strain.
%
%
We provided a phase diagram demonstrating that ordered oxygen vacancies (LaCoO$_{2.67}$) are the most probable explanation for the $3 \times 1$ pattern frequently observed in transmission electron microscopy images,
whereas the stoichiometric phase stabilizes under oxygen-rich growth conditions.
This allows to reconcile contradictory experimental results.
It also revealed that growth of epitaxial LaCoO$_{2.5}$ films in brownmillerite structure is impeded by preferential formation of competing oxides.

\begin{acknowledgments}

This work was supported by the German Research Foundation (Deutsche Forschungsgemeinschaft, DFG) within the SFB/TRR~80 (Projektnummer 107745057), Projects No.~G3 and No.~G8.
Computing time was granted by the Center for Computational Sciences and Simulation of the University of Duisburg-Essen
(DFG Grants No.~INST 20876/209-1 FUGG and No.~INST 20876/243-1 FUGG).

\end{acknowledgments}


%

\end{document}